\documentclass[pdftex,twocolumn,epjc3]{svjour3} 

\RequirePackage[T1]{fontenc}
\RequirePackage{graphicx}
\RequirePackage{flushend}
\RequirePackage[numbers,sort&compress]{natbib}
\RequirePackage[colorlinks,citecolor=blue,urlcolor=blue,linkcolor=blue]{hyperref}

\usepackage{tabularx}
\usepackage{graphicx}
\usepackage{dcolumn}
\usepackage{bm}
\usepackage{amssymb, amsmath, amsfonts}
\usepackage{numprint, color}
\usepackage[english]{babel}
\usepackage{newtxtext,newtxmath} 
\usepackage{float}
\usepackage{xcolor}
\usepackage[super,negative]{nth}

\newcommand{\mr}[1]{\mathrm{#1}}

\begin{document}

\title{Two-Year Optical Site Characterization for the Pacific Ocean Neutrino Experiment (P\babelhyphen{nobreak}ONE) in the Cascadia Basin
}

\author{
Nicolai Bailly\thanksref{addr2}\and
Jeannette Bedard\thanksref{addr2}\and
Michael B\"ohmer\thanksref{addr1}\and
Jeff Bosma\thanksref{addr2}\and
Dirk Brussow\thanksref{addr2}\and
Jonathan Cheng\thanksref{addr2}\and
Ken Clark\thanksref{addr3}\and
Beckey Croteau\thanksref{addr2}\and
Matthias Danninger\thanksref{addr4}\and
Fabio De Leo\thanksref{addr2}\and
Nathan Deis\thanksref{addr2}\and
Matthew Ens\thanksref{addr4}\and
Rowan Fox\thanksref{addr2}\and
Christian Fruck\thanksref{addr1}\and
Andreas G\"artner\thanksref{addr5}\and
Roman Gernh\"auser\thanksref{addr1}\and
{\color{black}Dilraj Ghuman\thanksref{addr4}\and}
Darren Grant\thanksref{addr6}\and
Helen He\thanksref{addr2}\and
Felix Henningsen\thanksref{addr7}\and
Kilian Holzapfel\thanksref{addr1}\and
Ryan Hotte\thanksref{addr2}\and
Reyna Jenkyns\thanksref{addr2}\and
Hamish Johnson\thanksref{addr4}\and
Akanksha Katil\thanksref{addr5}\and
Claudio Kopper\thanksref{addr6}\and
Carsten B. Krauss\thanksref{addr5}\and
Ian Kulin\thanksref{addr2}\and
Klaus Leism\"uller\thanksref{addr1}\and
Sally Leys\thanksref{addr8}\and
Tony Lin\thanksref{addr2}\and
Paul Macoun\thanksref{addr2}\and
{\color{black}Matthew Man\thanksref{addr5}\and}
Thomas McElroy\thanksref{addr5}\and
Stephan Meighen-Berger\thanksref{addr1}\and
Jan Michel\thanksref{addr9}\and
Roger Moore\thanksref{addr5}\and
Mike Morley\thanksref{addr2}\and
Laszlo Papp\thanksref{addr1}\and
Benoit Pirenne\thanksref{addr2}\and
Tom Qiu\thanksref{addr2}\and
Mark Rankin\thanksref{addr2}\and
Immacolata Carmen Rea\thanksref{addr1}\and
Elisa Resconi\thanksref{addr1}\and
Adrian Round\thanksref{addr2}\and
Albert Ruskey\thanksref{addr2}\and
Ryan Rutley\thanksref{addr2}\and
Christian Spannfellner\thanksref{addr1}\and
Jakub Stacho\thanksref{addr4}\and
Ross Timmerman\thanksref{addr2}\and
Meghan Tomlin\thanksref{addr2}\and
Matt Tradewell\thanksref{addr2}\and
Michael Traxler\thanksref{addr10}\and
Matt Uganecz\thanksref{addr2}\and
Seann Wagner\thanksref{addr2}\and
Juan Pablo Ya\~nez\thanksref{addr5}\and
Yinsong Zheng\thanksref{addr2}
}

\institute{
Department of Physics, Technical University of Munich, Garching, Germany\label{addr1}\and
Ocean Networks Canada, University of Victoria, Victoria, British Columbia, Canada\label{addr2}\and
Department of Physics, Engineering Physics and Astronomy, Queen’s University, Kingston, Ontario, Canada\label{addr3}\and
Department of Physics, Simon Fraser University, Burnaby, British Columbia, Canada\label{addr4}\and
Department of Physics, University of Alberta, Edmonton, Alberta, Canada\label{addr5}\and
Department of Physics and Astronomy, Michigan State University, East Lansing, MI, USA\label{addr6}\and
Max-Planck-Institut für Physik, Munich, Germany\label{addr7}\and
Department of Biological Sciences, University of Alberta, Edmonton, Alberta, Canada\label{addr8}\and
Institut für Kernphysik, Goethe Universität, Frankfurt, Germany\label{addr9}\and
Gesellschaft für Schwerionenforschung, Darmstadt, Germany\label{addr10}
}


\maketitle

\begin{abstract}
The STRings for Absorption length in Water (STRAW)
{\color{black}are the first in a series of pathfinders}
for the Pacific Ocean Neutrino Experiment (P-ONE), a future large-scale neutrino telescope in the north-eastern Pacific Ocean. STRAW consists of two $150\,\mr{m}$ long mooring lines instrumented with optical emitters and detectors. The
{\color{black}pathfinder}
is designed to measure the attenuation length of the water and perform a long-term assessment of the optical background at the future P-ONE site. After two years of continuous operation, measurements from STRAW show an optical attenuation length of about 28 metres at $450\,\mr{nm}$. Additionally, the data
{\color{black}allow}
a study of the ambient undersea background.
The overall optical environment reported here is comparable to other deep-water neutrino telescopes and qualifies the site for the deployment of P-ONE.

\end{abstract}

\section{Introduction}

\begin{figure*}[ht]
\centering
\includegraphics[width=0.8\textwidth]{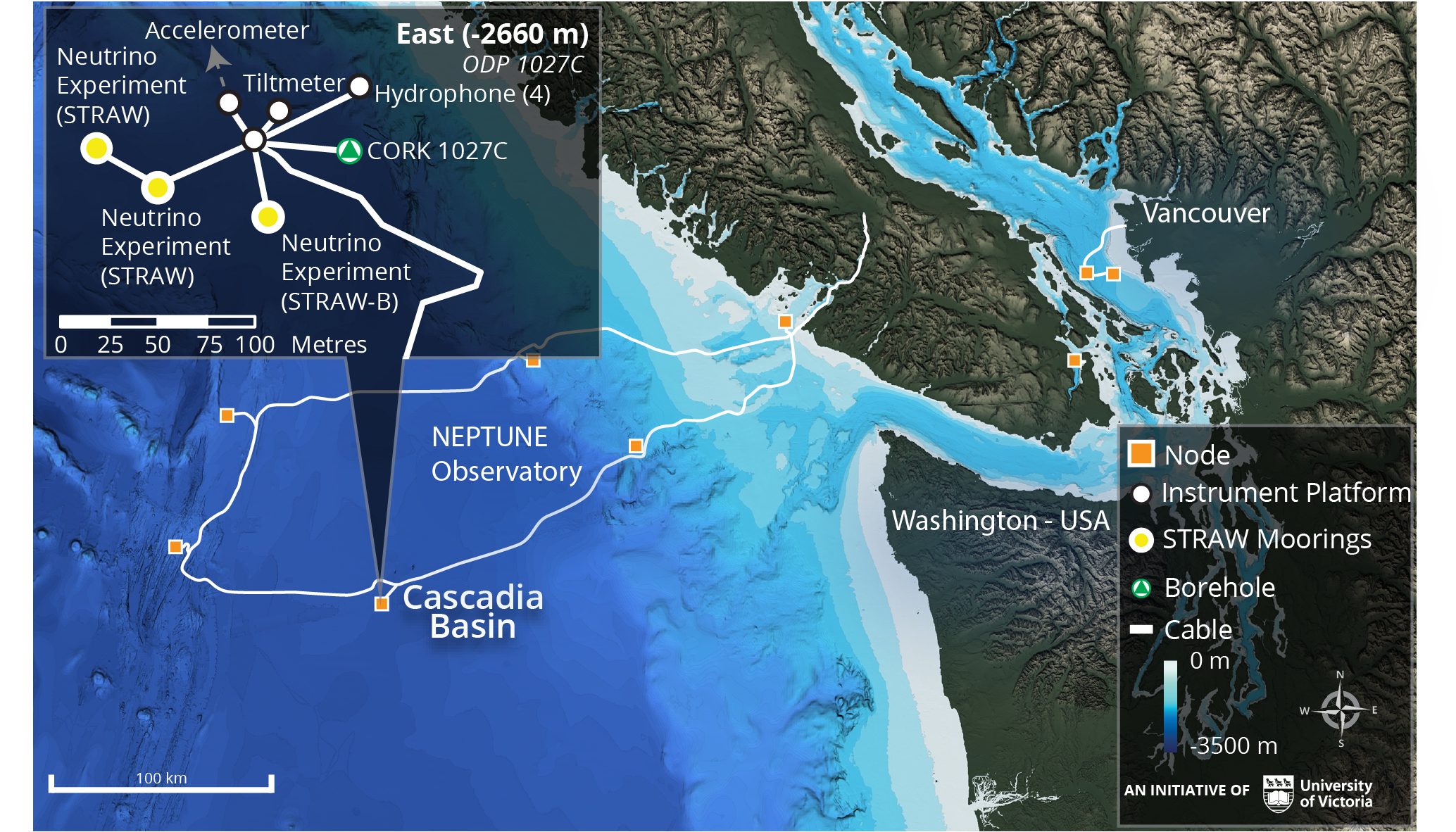}
\caption{Map of ONC's NEPTUNE Observatory. The two STRAW mooring lines are located in the Cascadia Basin at a depth of $2660\,\mr{m}$. The map also shows STRAW-b, the
{\color{black}successor}
to STRAW, which is currently taking data to complement the results reported here.}
\label{fig:map}
\end{figure*}

The Pacific Ocean Neutrino Experiment (P-ONE) is a proposed multi-cubic kilometre neutrino telescope deep in the Northern Pacific Ocean that will complement the sky coverage of other neutrino telescopes \cite{Agostini2020}. The development of P-ONE is possible thanks to Ocean Networks Canada (ONC),
a Canadian ocean research observing facility hosted at the University of Victoria,
that operates various undersea sensors and data transmission networks \cite{neptune}. An established ONC site, known as the Cascadia Basin (fig. \ref{fig:map}), 2,660$\,$m below sea level, provides an ideal environment for a large scale undersea neutrino telescope. As a precursor to P-ONE, a
{\color{black}pathfinder}
known as STRAW (STRings for Absorption length in Water) was developed and deployed in 2018 to measure the optical properties of the Cascadia Basin water as well as the ambient light levels \cite{Boehmer_2019}.

In this paper, the optical characterization of the Cascadia Basin is presented, based on two years of STRAW operation. First, the STRAW instrumentation and modes of operation are described. This is followed by an analysis of the optical deep-water environment. The optical attenuation length of the water at different wavelengths is extracted from the data and the ambient light levels from radioactive potassium decay and bioluminescence are discussed.
{\color{black}Both attenuation length and ambient light levels are fundamental characteristics of a neutrino detector site, as the attenuation length determines the module spacing of the neutrino telescope, and the ambient light levels dictate necessary DAQ and trigger capabilities.}

\section{Experiment Setup}
\label{setup}

\begin{figure}[ht]
\centering
\includegraphics[width=0.48\textwidth]{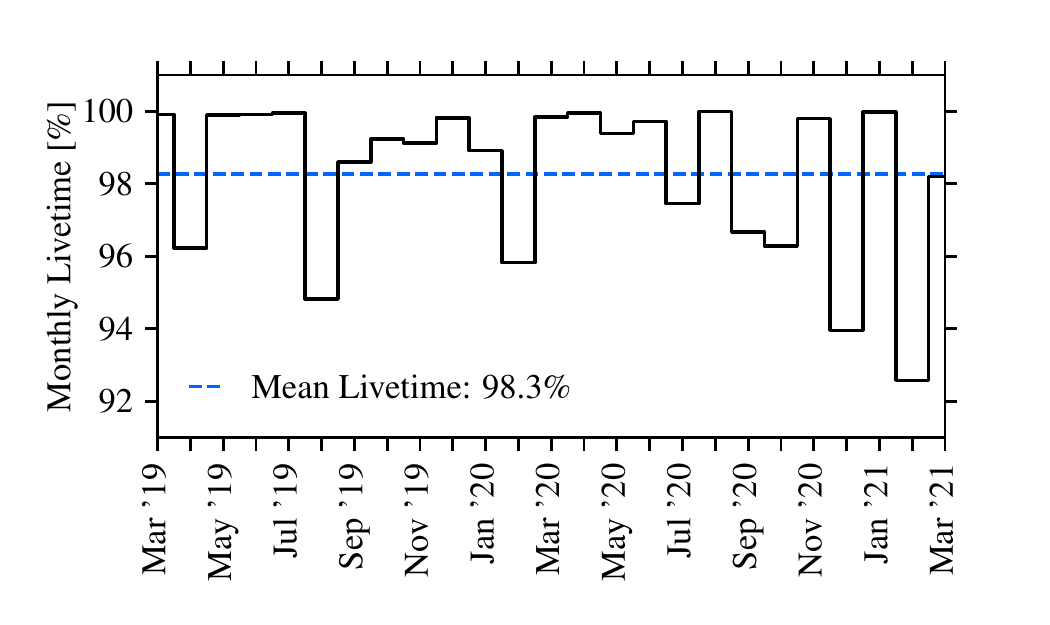}
\caption{Monthly livetime of STRAW over the last two years, during which a mean livetime of 98.3\% was achieved. The primary downtime periods are due to planned maintenance shutdowns of the undersea network.}
\label{fig:STRAW-up}
\end{figure}

The STRAW
{\color{black}pathfinder}
was deployed in June 2018 \cite{Boehmer_2019}. During a commissioning phase, test data of the individual modules
{\color{black}were}
taken and a DAQ system capable of simultaneous data taking and transfer to shore was developed. Continuous operation began in March 2019 and has been maintained since then, apart from short downtimes mostly caused by planned power outages and maintenance shutdowns of ONC's undersea network. Over the two years of operation, an average livetime of 98.3\% was achieved (fig. \ref{fig:STRAW-up}).

\begin{figure}[ht]
\centering
\includegraphics[width=0.48\textwidth]{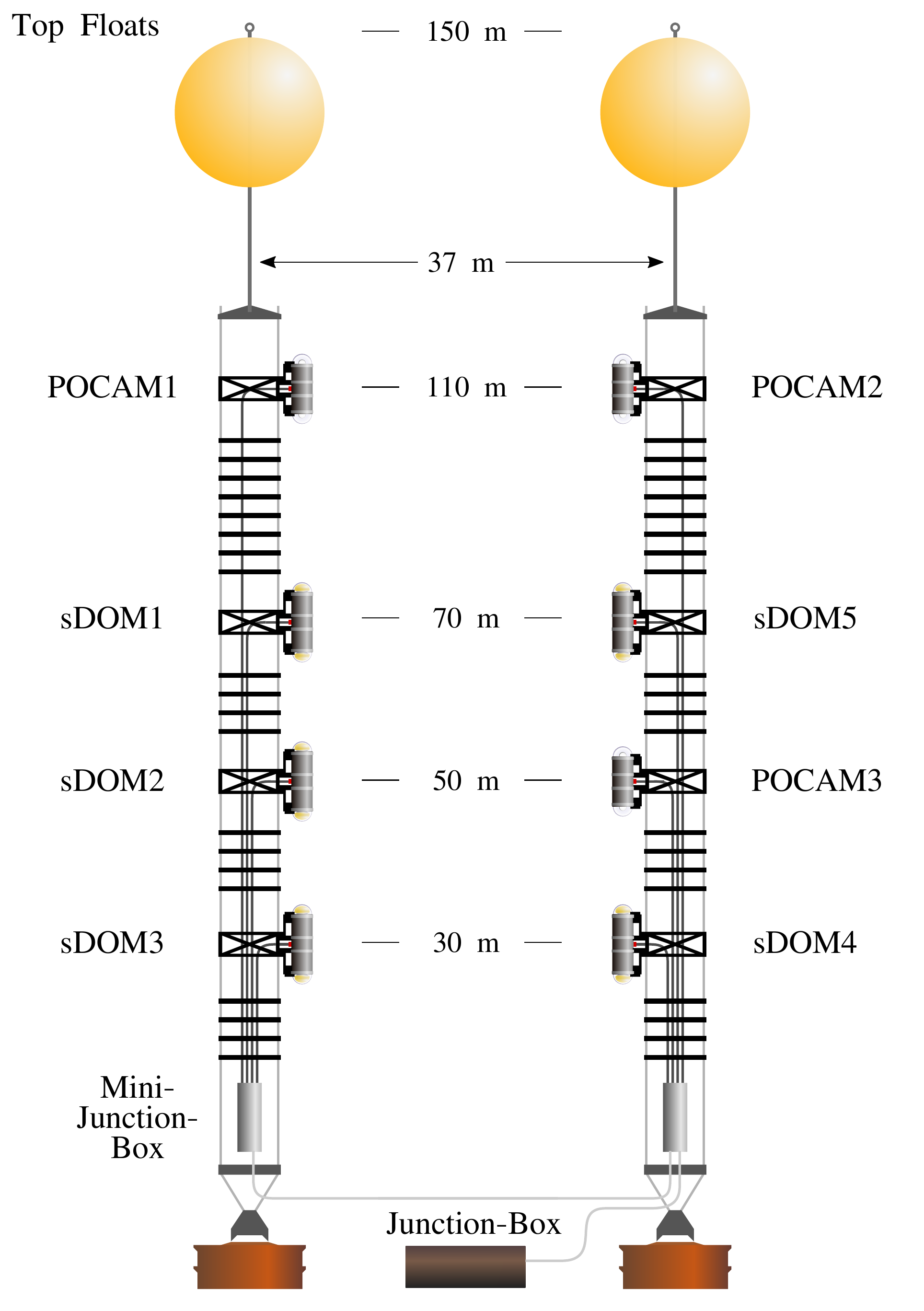}
\caption{Detailed technical sketch of the two STRAW mooring lines showing the position of all modules.}
\label{fig:STRAW-scheme}
\end{figure}

{\color{black}STRAW}
consists of two 150$\,$m long mooring lines spaced 37$\,$m apart
{\color{black}(fig. \ref{fig:STRAW-scheme})}.
Different optical modules are mounted along each mooring line at 30$\,$m, 50$\,$m, 70$\,$m, and 110$\,$m above the seafloor. The three POCAMs (Precision Optical CAlibration Modules)\cite{Henningsen:2020zsj} act as light emitters, the five sDOMs (STRAW Digital Optical Modules) act as optical receivers.
All modules are housed in titanium cylinders with glass hemispheres at both ends, one hemisphere facing upwards and the other facing downwards.

A POCAM is equipped with an LED emitter in each hemisphere. A Kapustinsky driver circuit \cite{KAPUSTINSKY1985612} creates flashes of
{\color{black}four to eight}
nanoseconds in length with an adjustable intensity
{\color{black}\cite{Boehmer_2019}}.
The LEDs are placed behind an integrating sphere to create nearly isotropic light flashes.
Four different LEDs are available, allowing a measurement of the attenuation length at four different wavelengths: 365$\,$nm, 400$\,$nm, 450$\,$nm and 585$\,$nm. It should be noted that these are not the nominal wavelengths given by the manufacturers. The high voltages that the Kapustinsky circuit uses to drive the LEDs change their emission spectrum and the transmission curve of the water changes the peak wavelength at which the attenuation can be probed.
{\color{black}The wavelengths reported in this paper are taking these effects into account.}

An sDOM houses a Hamamatsu Photonics R12199 photomultiplier tube (PMT) in each hemisphere. The PMTs are read out by a time-to-digital converter that uses the Trigger Readout Board (TRB3sc) developed by the German heavy ion research centre GSI \cite{TRB}. This readout system allows the detectors to run in two different modes of operation. In the default low-precision mode, the sDOMs count the number of pulses detected in a time interval of $30\,\mr{ms}$. This mode was used to measure the ambient background over two years. In high-precision mode, the exact timestamp of each pulse is stored with sub-nanosecond precision
{\color{black}relative to the master clock in one of the mini junction boxes},
allowing an analysis on the single-photon level. Due to the high data rate, the sDOMs can only be operated for a few minutes at a time in high-precision mode before the front-end buffers fill up.

\section{Attenuation Measurement}
\label{optics_intro}

The primary analysis of this paper is concerned with the optical attenuation length of the seawater which takes both scattering and absorption into account. The absorption length $l_\mr{abs}$ is the distance after which the probability of a photon not being absorbed is $1/e$.
Equally, the scattering length $l_\mr{scat}$
{\color{black}(sometimes called geometric scattering length)}
is the distance after which the probability of a photon not being scattered is $1/e$ where scattering is defined as a change in the photon's direction.

Light is emitted as an
{\color{black} almost}
isotropic flash from the POCAMs with intensity $I_0$. Travelling through the water, absorption and scattering will reduce its intensity. The intensity $I$ of direct (unscattered) light at a distance $r$ can be modelled using an exponential law
\begin{equation}
    I(r)=\frac{I_0}{4\pi r^2} \cdot\exp\left(-\frac{r}{l_\mr{abs}}-\frac{r}{l_\mr{scat}}\right)=\frac{I_0}{4\pi r^2}\cdot\exp\left(-\frac{r}{l_\mr{att}}\right),
\end{equation}
with the attenuation length $l_\mr{att}$ defined as
\begin{equation}
    l_\mr{att}=\left(l_\mr{abs}^{-1}+l_\mr{scat}^{-1}\right)^{-1}.
\end{equation}
While scattered light will add to the intensity measured at a distance, it will take a longer path than direct light and therefore arrive later. Timing information can be used to filter out scattered light so that its effect on the attenuation length measurement is
{\color{black}small. As our timing is not perfect (see section \ref{method}) some scattered light will enter our measurement. Previous measurements by ANTARES \cite{Antares} and Baikal-GVD \cite{baikal_scatter} have measured scattering as much weaker than absorption in water. The absorption length is therefore expected to be the dominant contribution to the attenuation length.}

It should be noted that different experiments use slightly different definitions of the attenuation length, which must be taken into account when making comparisons.

\subsection{Method}\label{method}

The measurement of the attenuation length is based on the
{\color{black}hit fraction}
$h$, the number of events per POCAM flash that are detected by an sDOM, which is extracted from STRAW data and compared to the predictions of a parametric model.

To extract $h$, the POCAMs are flashed at a fixed interval $T$. As the sDOMs record timestamps $t_i$ of the rising edges of all events, a histogram of $t_i\,\mr{modulo}\,T$ (phase) shows a clear peak for the POCAM signal. Background events are equally distributed over all phases. To correct for clock drift, $T$ is modelled as a second order polynomial in time with very small first and second order coefficients. The coefficients are adjusted to maximize the peak height of the phase histogram, as this corresponds to the sharpest POCAM signal.

\begin{figure*}[htb]
\centering
\includegraphics[width=0.98\textwidth]{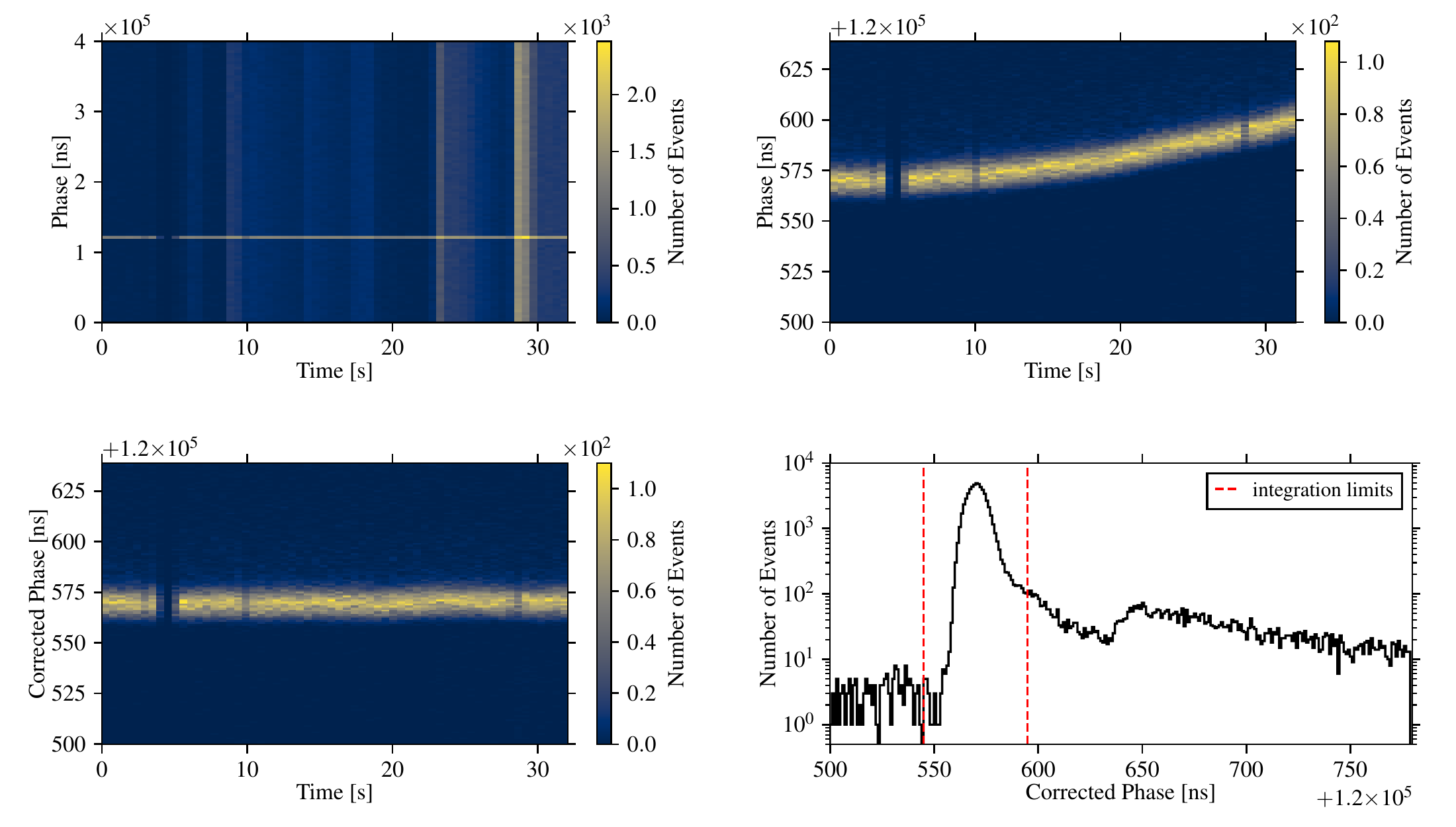}
\caption{A visual representation of the signal extraction process using 32 seconds of data from a single PMT. \textbf{Top left:} Plotting the phase (event time modulus POCAM interval $T$) over time shows the POCAM signal as a horizontal line at a phase of approximately
${\color{black}y=1.2\cdot10^5\mr{ns}}$.
The background rate changes slowly over time due to bioluminescence, resulting in vertical lines. \textbf{Top right:} Zooming in
{\color{black}on the y-axis at $y=1.2\cdot10^5\mr{ns}$ and using a finer binning},
the effect of the clock drift becomes visible. Gaps can be seen where data exceeding the DAQ capabilites were discarded. \textbf{Bottom left:} After adding small first- and second-order corrections to $T$, the POCAM signal has
{\color{black}an almost} constant
phase throughout the entire measurement.
{\color{black}A small variation can still be seen, as the description of $T$ as a second order function is only an approximation.}
\textbf{Bottom right:} The histogram of the phases over the entire time period is integrated from $-25\,\mr{ns}$ to $+25\,\mr{ns}$ around its peak. A long tail of scattered light can be seen, with a short drop after the peak due to the DAQ dead-time.
{\color{black} The FWHM of the peak with $12\,\mr{ns}$ is close to the sum of the POCAM pulse FWHM of $6\,\mr{ns}$ and the PMT transit time spread with a FWHM of $3\,\mr{ns}$. An additional spread is caused by the small discrepancy between the real clock drift and the second-order approximation of $T$.}
}
\label{fig:sig-extract}
\end{figure*}

{\color{black} For strong signals, the FWHM of the peak is in the order of $8-14\,\mr{ns}$ and is in reasonable agreement with the expected value based on the POCAM pulse FWHM of $5-8\,\mr{ns}$ and the PMT transit time spread of $2.5-4\,\mr{ns}$, depending on the operating voltage \cite{Boehmer_2019}. For weaker signals, the determination of $T$ is not ideal and leads to a signal spread over up to $50\,\mr{ns}$. Therefore a $50\,\mr{ns}$ integration window centered at the peak was chosen (fig. \ref{fig:sig-extract}). The background is subtracted from the phase histogram before integration.}

The sDOMs take time-over-threshold measurements, with the threshold set to half of the average single-photon response of the PMT. The clocks of all sDOMs are synchronized, allowing the detection of very weak signals in far away sDOMs by using sDOMs closer to the POCAM as timing reference.

Three effects stemming from detector electronics have to be taken into account for this analysis. The first is a straight-forward $70\,\mr{ns}$ dead-time after each detected pulse, caused by both the dead-time of the DAQ and a reduced detection efficiency until the capacitors in the signal filters are recharged. This effect is reduced by using only data with $h<1$. A second type of dead-time is caused by high bioluminescence rates that exceed the capabilities of the DAQ.
{\color{black}These}
data
{\color{black}were}
discarded from the attenuation length analysis. Third, there are rare periods of a few seconds where only one of the two PMTs in an sDOM detector sees a signal. This is most likely caused by a short breakdown of the high-voltage supplied to the PMT during periods of high bioluminescence activity. In this case, the data
{\color{black}were}
also discarded.

To test the stability of the signal extraction method, the POCAMs were operated at the same settings over several hours. The extracted $h$ was monitored during the entire run time and found to be stable within uncertainties for most POCAM-sDOM combinations. It was found that $h$ is unstable when an sDOM lies in the shadow of another module. Since the entire detector assembly moves slightly in the water current, the impact of shadows on $h$ cannot be predicted and the affected POCAM-sDOM combinations are excluded from the analysis. While this was expected, an unstable signal for the combination POCAM2-sDOM5 was also found. It is suspected that a data cable with too much slack between the modules is casting a shadow and so this specific combination is excluded. The attenuation length analysis was therefore restricted to sDOMs 1, 4, and 5 for POCAM1, and sDOMs 1, 2, and 3 for POCAM2 (see fig. \ref{fig:STRAW-scheme} for reference). POCAM3 is not used in the analysis.

A large data set was taken over several days in autumn 2020, consisting of 96 individual runs. One run includes of a full series of measurements for one wavelength at five different intensities for two POCAMs. For each intensity, each POCAM flashed for about 30$\,$s at 2.5$\,$kHz. A single run is sufficient to fit the attenuation length for a specific wavelength. To test the stability of the results, 24 runs were performed at each wavelength. Additionally, a measurement at $450\,\mr{nm}$ with a low flasher intensity was added to each run, providing a common baseline.

{\color{black}These}
data
{\color{black}are}
complemented by runs taken regularly over two years of STRAW operation.

\subsection{Model description}\label{sec:modelfit}

A parametric model of the entire STRAW detector was constructed to measure the attenuation length $l_{\mr{att}}$. Accounting for the emission, propagation, and reception of light, the model predicts $h$ for all sDOMs and can be fitted to the measurements. Most of the parameters, like calibration constants and parameters describing the geometry, are nuisance parameters that can only be measured with some degree of uncertainty and are constrained by a prior. The only parameter that is unconstrained in the fit is $l_{\mr{att}}$. Combining a large number of measurements using a variety of POCAM/sDOM combinations with different baselines and different flash intensities makes it possible to constrain the nuisance parameters.

The {\color{black}Poisson mean} number of photons  detected by an sDOM is modeled as
\begin{equation}
    {\color{black}\mu} = \frac{p \cdot s}{4\pi r^2} \cdot \exp\left(-\frac{r}{l_\mr{att}}\right)
\end{equation}
where $p$ includes all POCAM parameters, $s$ all sDOM parameters, and $r$ is the distance between sDOM and POCAM.
{\color{black} The angular profiles of the modules are described by functional approximations of the measurements published in a previous paper \cite{Boehmer_2019}.}
The function $p$ is defined as
\begin{equation}
    p = N_0 \cdot p_\mr{rel} \cdot \left[ 0.75 + 0.25 \cdot \left|\,\cos(\theta)^\frac{2}{3} \right| \right]
\end{equation}
with the number of emitted photons $N_0$ based on lab measurements,  a correction factor $p_\mr{rel}$, and the angle $\theta$ of the direct line from POCAM to sDOM relative to the vertical. The function $s$ is defined as
\begin{equation}
    s = A_\mr{eff} \cdot s_\mr{rel} \cdot \eta \cdot \epsilon \cdot \left| \, \cos\left( \frac{\theta}{2} \right)^{\gamma} \right|
\end{equation}
with the effective area $A_\mr{eff}$ of the PMT, a correction factor $s_\mr{rel}$, the quantum efficiency $\eta$ of the PMT, the probability $\epsilon$ that a PMT signal triggers the DAQ, and a geometric factor $\gamma$ describing the angular detection profile of the sDOMs. The variables $A_\mr{eff}$ and $N_0$ are treated as fixed parameters and are not fitted. Different $N_0$, $p_\mr{rel}$, and $\eta$ are implemented for each wavelength, equally different $p_\mr{rel}$ and $s_\mr{rel}$ are implemented for each module. Additionally, the model allows for a small vertical offset $y_\mr{off}$ between the two mooring lines which results in different $\theta$ and $r$. A full list of the parameters is given in table \ref{tab:parameters}.

\begin{table*}
  \begin{center}
    \color{black}
    \caption{Parameters for the attenuation length fit and results of the Markov Chain Monte Carlo sampling. Priors are given with Gaussian sigma. Results extracted from posterior distributions correspond to the \nth{50}, \nth{16} and \nth{84} percentile, respectively.}
    \label{tab:parameters}
    \begin{tabularx}{0.7\textwidth}{p{0.1\textwidth} p{0.3\textwidth} p{0.1\textwidth} p{0.1\textwidth}}
    \hline\hline
    Parameter & Description   & Prior      & Result \\
    \hline
    $l_\mr{att}(365\,\mr{nm})$      & attenuation length at $365\,\mr{nm}$       & none                          & $10.4\substack{+0.4\\[-0.07em]-0.3}\,\mr{m}$ \\
    $l_\mr{att}(400\,\mr{nm})$      & attenuation length at $400\,\mr{nm}$       & none                          & $14.6\substack{+0.4\\[-0.07em]-0.6}\,\mr{m}$ \\
    $l_\mr{att}(450\,\mr{nm})$      & attenuation length at $450\,\mr{nm}$      & none                          & $27.7\substack{+1.9\\[-0.07em]-1.3}\,\mr{m}$ \\
    $l_\mr{att}(585\,\mr{nm})$      & attenuation length at $585\,\mr{nm}$       & none                          & $7.1\substack{+0.4\\[-0.07em]-0.3}\,\mr{m}$  \\
    $\eta(365\,\mr{nm})$            & PMT quantum efficiency at $365\,\mr{nm}$          & $0.24 \pm 0.02$               & $0.22\substack{+0.02\\[-0.07em]-0.02}$ \\
    $\eta(400\,\mr{nm})$            & PMT quantum efficiency at $400\,\mr{nm}$          & $0.24 \pm 0.02$               & $0.24\substack{+0.02\\[-0.07em]-0.02}$  \\
    $\eta(450\,\mr{nm})$            & PMT quantum efficiency at $450\,\mr{nm}$          & $0.21 \pm 0.02$               & $0.24\substack{+0.02\\[-0.07em]-0.02}$  \\
    $\eta(585\,\mr{nm})$            & PMT quantum efficiency at $585\,\mr{nm}$          & $0.04 \pm 0.01$               & $0.04\substack{+0.01\\[-0.07em]-0.01}$ \\
    $\epsilon$                      & sDOM trigger efficiency & $0.75  \pm 0.25$              & $0.83\substack{+0.10\\[-0.07em]-0.10}$  \\
    $\gamma$                        & sDOM angular profile coefficient & $4.0\, \pm 0.2$               & $3.98\substack{+0.10\\[-0.07em]-0.09}$  \\
    $y_{\mr{off}}$                  & vertical offset between strings  & $0.0\,\mr{m} \pm 1.0\,\mr{m}$ & $0.08\substack{+0.44\\[-0.07em]-0.44}\,\mr{m}$  \\
    $p_\mr{1, rel}(365\,\mr{nm})$   & POCAM1 correction factor at $365\,\mr{nm}$    & $1.0 \pm 0.1$                 & $0.77\substack{+0.06\\[-0.07em]-0.06}$  \\
    $p_\mr{2, rel}(365\,\mr{nm})$   & POCAM2 correction factor at $365\,\mr{nm}$   & $1.0 \pm 0.1$                 & $1.07\substack{+0.09\\[-0.07em]-0.09}$  \\
    $p_\mr{1, rel}(400\,\mr{nm})$   & POCAM1 correction factor at $400\,\mr{nm}$    & $1.0 \pm 0.1$                 & $0.99\substack{+0.07\\[-0.07em]-0.06}$  \\
    $p_\mr{2, rel}(400\,\mr{nm})$   & POCAM2 correction factor at $400\,\mr{nm}$    & $1.0 \pm 0.1$                 & $1.00\substack{+0.08\\[-0.07em]-0.08}$  \\
    $p_\mr{1, rel}(450\,\mr{nm})$   & POCAM1 correction factor at $450\,\mr{nm}$    & $1.0 \pm 0.1$                 & $1.10\substack{+0.07\\[-0.07em]-0.07}$  \\
    $p_\mr{2, rel}(450\,\mr{nm})$   & POCAM2 correction factor at $450\,\mr{nm}$    & $1.0 \pm 0.1$                 & $1.08\substack{+0.07\\[-0.07em]-0.07}$  \\
    $p_\mr{1, rel}(585\,\mr{nm})$   & POCAM1 correction factor at $585\,\mr{nm}$    & $1.0 \pm 0.1$                 & $0.99\substack{+0.10\\[-0.07em]-0.10}$  \\
    $p_\mr{2, rel}(585\,\mr{nm})$   & POCAM2 correction factor at $585\,\mr{nm}$    & $1.0 \pm 0.1$                 & $1.00\substack{+0.10\\[-0.07em]-0.10}$  \\
    $s_\mr{1, rel}$                 & sDOM1 correction factor     & $1.0 \pm 0.25$                & $1.17\substack{+0.07\\[-0.07em]-0.07}$  \\
    $s_\mr{2, rel}$                 & sDOM2 correction factor     & $1.0 \pm 0.25$                & $0.99\substack{+0.07\\[-0.07em]-0.07}$  \\
    $s_\mr{3, rel}$                 & sDOM3 correction factor     & $1.0 \pm 0.25$                & $0.88\substack{+0.07\\[-0.07em]-0.08}$  \\
    $s_\mr{4, rel}$                 & sDOM4 correction factor     & $1.0 \pm 0.25$                & $0.86\substack{+0.09\\[-0.07em]-0.09}$  \\
    $s_\mr{5, rel}$                 & sDOM5 correction factor     & $1.0 \pm 0.25$                & $1.14\substack{+0.07\\[-0.07em]-0.07}$  \\
    \hline\hline
    \end{tabularx}
  \end{center}
\end{table*}

The above describes the model for predicting the average number of photons detected for a given POCAM/sDOM combination. Poissonian statistics are used to compare this to the measured quantity $h$
\begin{equation}
    h = 1 - P_{\color{black}\mu}(0) = {\color{black}1 - e^{-\mu}}\, .
\end{equation}
A simple Gaussian likelihood is used for the fit
\begin{equation}
    \log( \mathfrak{L} ) = \sum_{i}{-\frac{(h_i - h_{i,\mr{model}})^2}{\Delta h_i^2}}\, ,
\end{equation}
where $i$ runs over the individual
{\color{black}hit fraction}
measurements $h_i$ with uncertainties $\Delta h_i$.

The probability distribution is sampled with the Markov-Chain Monte Carlo (MCMC) method using \emph{emcee} \cite{foreman-mackey_emcee_2013}. Several measurements with different baselines, flasher intensities, and wavelengths were combined. A Gaussian prior is used for all nuisance parameters, with the standard deviations given in table \ref{tab:parameters}.

\subsection{Cross-Check via Geant4 Simulation} \label{sec:propertiesG4}

As a cross-check to the method described in the previous section, Geant4 \cite{Agostinelli2003} was used to simulate the STRAW setup. The simulated Geant4 geometry itself was not modeled precisely after the detector. To improve performance, all sDOMs were simplified as spheres, and hundreds of sDOMs were placed in the simulation volume to increase statistics. Special care was taken to make sure that the simulated sDOMs did not cast shadows on each other. The POCAMs were simulated as perfectly isotropic point sources.

The actual angular emission profile of the POCAM and the actual angular detection profile of the sDOM were applied by reweighting the simulation results. Equally, only one long absorption length ($l_\mr{att}=60\,\mr{m}$) was simulated and the results were reweighted based on the total light path of each simulated photon.

The previously described model fit takes into account many detector parameters (table \ref{tab:parameters}) and uses Bayesian priors for these parameters, based on data sheets and lab measurements. The Geant4 simulation, on the other hand, does not use these priors. Specifically, only the angular detection/emission profiles of the modules and the geometry of the mooring lines were taken as fixed, all other parameters, such as flasher intensity, detection efficiency, and attenuation length, were fitted freely. Therefore, the Geant4 fit is less precise than the model fit of the previous section but less susceptible to errors in the detector model.

The Geant4 fit was used to thoroughly scrutinize the model fit, using both simulated and real data, and to cross-check the results presented in this paper.

\subsection{Results}

\begin{figure*}[ht]
\centering
\includegraphics[width=0.75\textwidth]{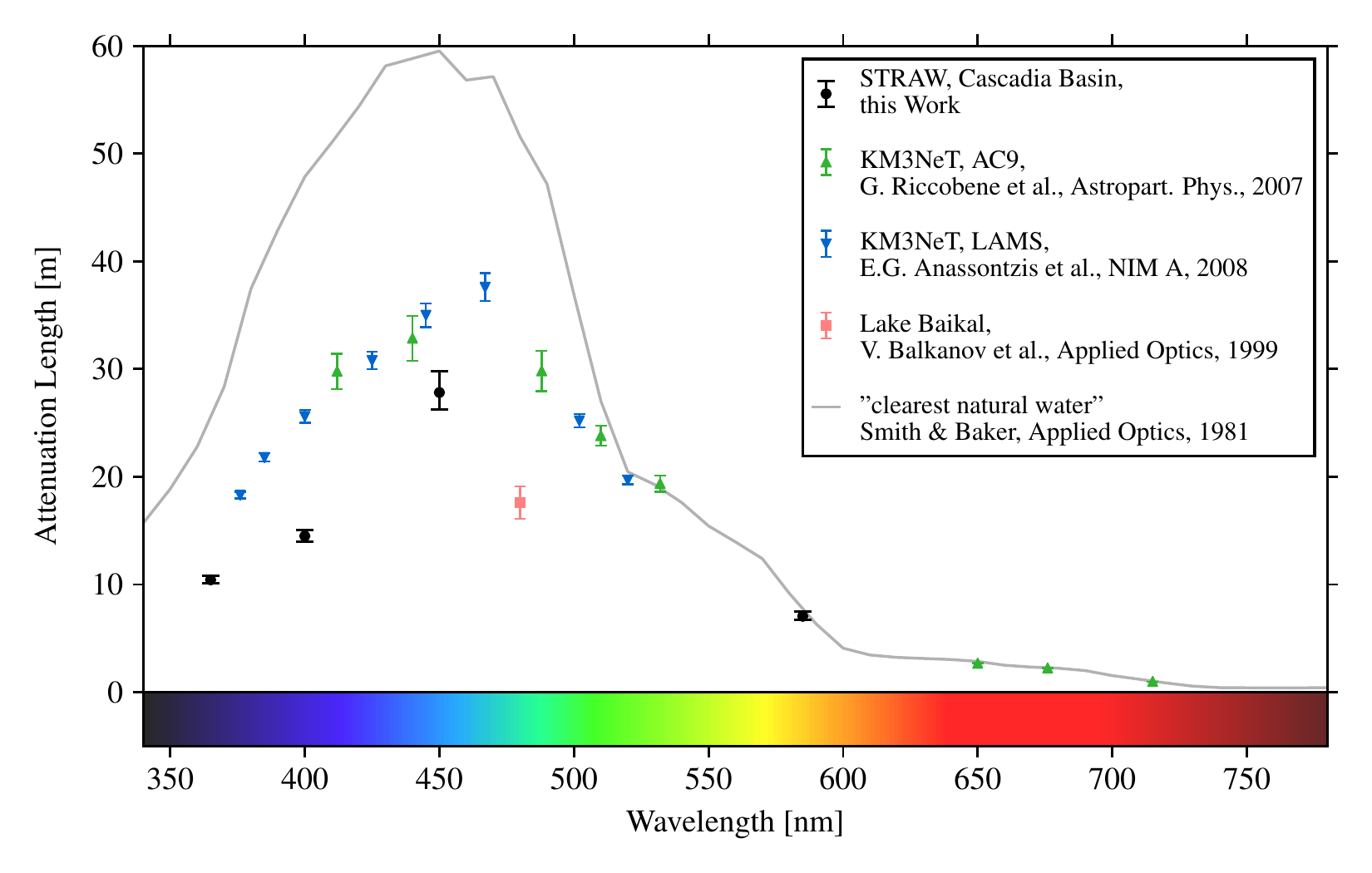}
\caption{
Attenuation length from a joint multi-wavelength fit of all available STRAW data. Measurements from the sites of other neutrino telescopes are plotted for comparison \cite{balkanov_situ_1999,riccobene_deep_2007,anassontzis_light_2011}, as well as an estimate of the attenuation length in the clearest ocean waters \cite{smith_optical_1981}. The attenuation length at Lake Baikal has been measured using the neutrino detector itself and an external flasher as a light source. The measurements
{\color{black}denoted KM3NeT} have been conducted {\color{black}in the Mediterranean sea}
with a WETLabs AC9 trasmissometer (green data points) and the Long Arm Marine Spectrophotometer (LAMS, blue data points) using LEDs and photodiodes. From the data measured with the AC9 the values obtained at 3000~m.b.s.l. and the site denoted PC1 have been plotted for reference. The LAMS values that have all been measured close to the Capo Passero site of KM3NeT have simply been averaged for this plot, while the standard deviation between the measurements is reflected in the error bars.
}
\label{fig:attenuation_wl}
\end{figure*}

\begin{figure}[ht]
\centering
\includegraphics[width=\columnwidth]{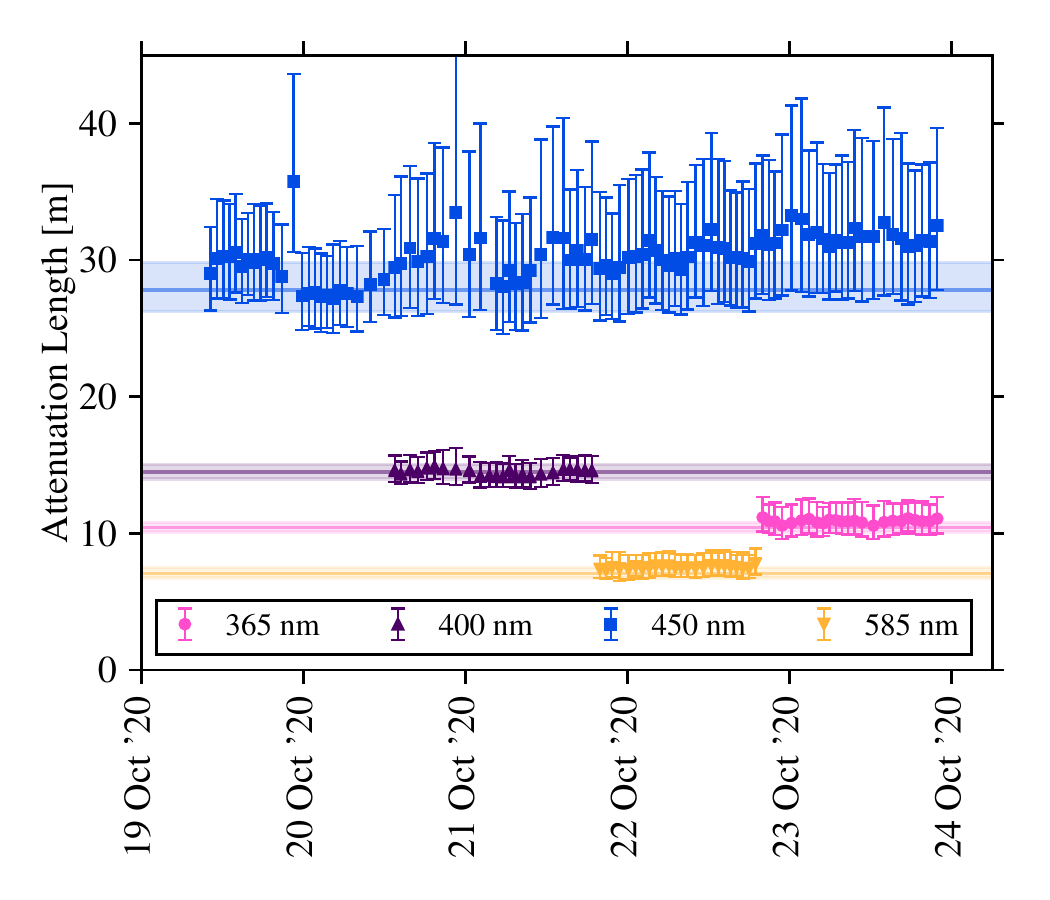}
\caption{Measured attenuation length over four days. 24 runs cycling through different flasher intensities have been recorded for each wavelength. Additionally, data
{\color{black}were}
taken at $450\,\mr{nm}$ with a single flasher intensity during the runs at other wavelengths, resulting in larger error bars. Each run has been fitted individually. The error bars on the individual runs also reflect the uncertainties of the nuisance parameters and are therefore highly correlated. The solid lines with error bands represent the results from the all-data joint fit (fig. \ref{fig:attenuation_wl}).}
\label{fig:attenuation_time}
\end{figure}

\begin{figure}[ht]
\centering
\includegraphics[width=\columnwidth]{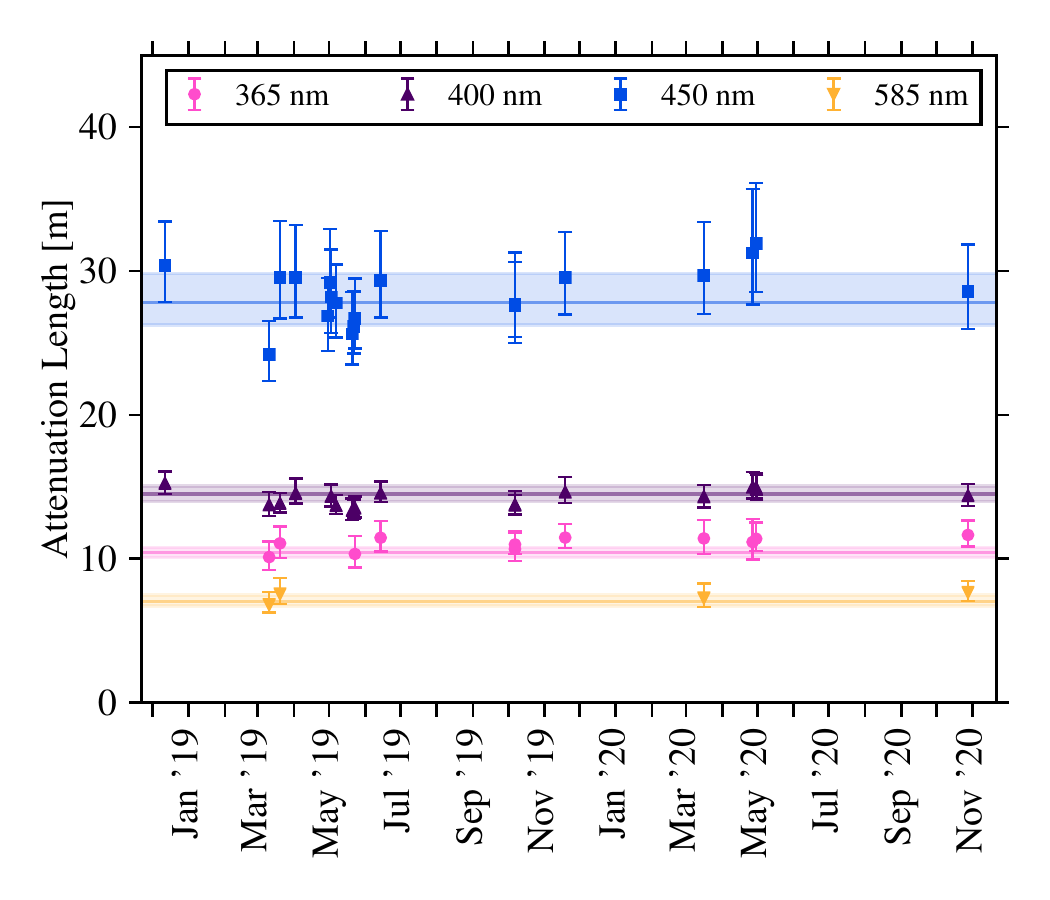}
\caption{Long term study of the attenuation length over two years. The error bars on the individual data points contain the uncertainties of the nuisance parameters and are therefore highly correlated. The solid lines with error bands represent the results from the all-data joint fit (fig. \ref{fig:attenuation_wl}). The data points from fig. \ref{fig:attenuation_time} are not shown in this plot.}
\label{fig:attenuation_time_archival}
\end{figure}

For the full data set, fitting the model to all measurements with the four different wavelengths simultaneously, the
attenuation lengths shown in table \ref{tab:attenuation_len} are obtained.
These measurements are also visualized in fig. \ref{fig:attenuation_wl} where they are compared to similar measurements conducted in the Mediterranean Sea {\color{black}(KM3Net site)} \cite{riccobene_deep_2007,anassontzis_light_2011}, in Lake Baikal \cite{balkanov_situ_1999}, and in the in the clearest ocean waters as reported by \cite{smith_optical_1981}.
{\color{black}
The measurements for the KM3NeT site have been conducted with a WETLabs AC9 transmissometer, which uses a collimated geometry and filters out scattered light, and the Long Arm Marine Spectrophotometer (LAMS) that uses a combination of LEDs and photodiodes and does not filter out scattered light. Nonetheless, the results are very similar, which agrees with the expectation that in water scattering is only a small contribution to the attenuation length.

To test the influence of scattered light in the STRAW measurements, the $450\mr{nm}$ fit was repeated using various larger integration windows. Even with an integration window of $400\,\mr{ns}$ which should include a significant amount of scattered light, the reconstructed attenuation length was only $10\%$ higher. It is therefore safe to assume that the measurements in fig. 5 are comparable despite the different methods used.
}

The results from the short-term variability study conducted in autumn 2020 are shown in fig. \ref{fig:attenuation_time}. The data points are the result of individually fitting data from a single run containing only measurements with one wavelength. The error bars are dominated by uncertainties of the nuisance parameters and are therefore highly correlated.
At $585\,\mr{nm}$ it was not always possible to extract the flasher signal from the background, as the short absorption length results in a very weak signal.
The $450\,\mr{nm}$ measurements accompanying runs 25-96
{\color{black}(Oct 20 - Oct 24)}
were only performed at a single flasher intensity, and therefore have larger uncertainties than the dedicated $450\,\mr{nm}$ measurements in runs 1-24
{\color{black}(Oct 19 - Oct 20)}.
Additionally, these measurements show a higher average attenuation length than the dedicated $450\,\mr{nm}$ measurements. This could be the result of a slight dependence of the LED spectrum on the flasher intensity or a higher transparency of the water column further away from the seafloor, as a measurement at a lower intensity favours the shorter baselines between the POCAMs and the uppermost sDOMs.

Some variation over time can be seen in the 450$\,$nm runs. This could potentially be explained by a slight tilt of the STRAW strings or a change in water composition, both of which could be caused by tidal currents. The monitoring of the attenuation length over two years is shown in fig. \ref{fig:attenuation_time_archival} and does not show any significant seasonal variability.

\begin{table}[ht]
  \begin{center}
    \caption{Results of the combined attenuation-length model fit , using all available data (central column, see also fig. \ref{fig:attenuation_wl}), compared to the Geant4 cross-check (right column).}
    \label{tab:attenuation_len}
    \begin{tabularx}{.48\textwidth}{l l l }
    \hline\hline
    Effective Central & Attenuation length & Attenuation length\\
    Wavelength & using model fit & using Geant4 cross-check\\
    \hline
    $365\,\mr{nm}$ & $10.4\substack{+0.4\\-0.3}\,\mr{m}$ & $12.4\pm2.6\,\mr{m}$\\[0.3em]
    $400\,\mr{nm}$ & $14.6\substack{+0.4\\-0.6}\,\mr{m}$ & $16.1\pm2.2\,\mr{m}$\\[0.3em]
    $450\,\mr{nm}$ & $27.7\substack{+1.9\\-1.3}\,\mr{m}$ & $29.4\pm3.5\,\mr{m}$\\[0.3em]
    $585\,\mr{nm}$ & $7.1\substack{+0.4\\-0.3}\,\mr{m}$ & $9.3\pm3.2\,\mr{m}$\\
    \hline\hline
    \end{tabularx}
  \end{center}
\end{table}

\section{Salinity and K-40 Measurement}

In addition to measuring the attenuation length of the Cascadia Basin seawater, STRAW measured the ambient light background present in the deep ocean which is essential for the future P-ONE trigger development. While the background consists of many stochastic spikes from bioluminescence, there is a continuous noise floor due to the decay of radioactive isotopes occurring in sea salt. The most prevalent contributor to the light background is potassium-40 ($^{40}$K) which has two main decay channels \cite{CHEN20171}:

\begin{align}
    & {^{40}\mr{K} \rightarrow\, ^{40}\mr{Ca} + \mr{e}^- + \overline{\nu}_{\mr{e}}}\label{eq:betaminus} & 89.3\%\\
    & {^{40}\mr{K} + \mr{e}^- \rightarrow\, ^{40}\mr{Ar} + \nu_{\mr{e}} + \gamma}\label{eq:ecap} & 10.7\%
\end{align}

The first corresponds to $\beta^-$ decay and produces electrons with energies up to 1.3 MeV which emit Cherenkov photons. The second corresponds to electron capture where the photon released by the excited argon nucleus can generate energetic electrons through Compton scattering, producing Cherenkov photons. Cherenkov light in both these decays shows up in the sDOM optical receivers as ambient background. This background is compared to a Geant4 simulation of the sDOM response to $^{40}$K.

\subsection{Method}

To isolate the $^{40}$K activity in STRAW data, only the photons coincident between the top and bottom PMTs of an sDOM within a $\Delta t<25\,\mr{ns}$ window are considered. Most coincidences occur randomly, resulting in a constant rate distribution in $\Delta t$. On the other hand, photons produced from the same $^{40}$K decay will pile up around $\Delta t=0$.

\begin{figure}[ht]
\centering
\includegraphics[width=0.48\textwidth]{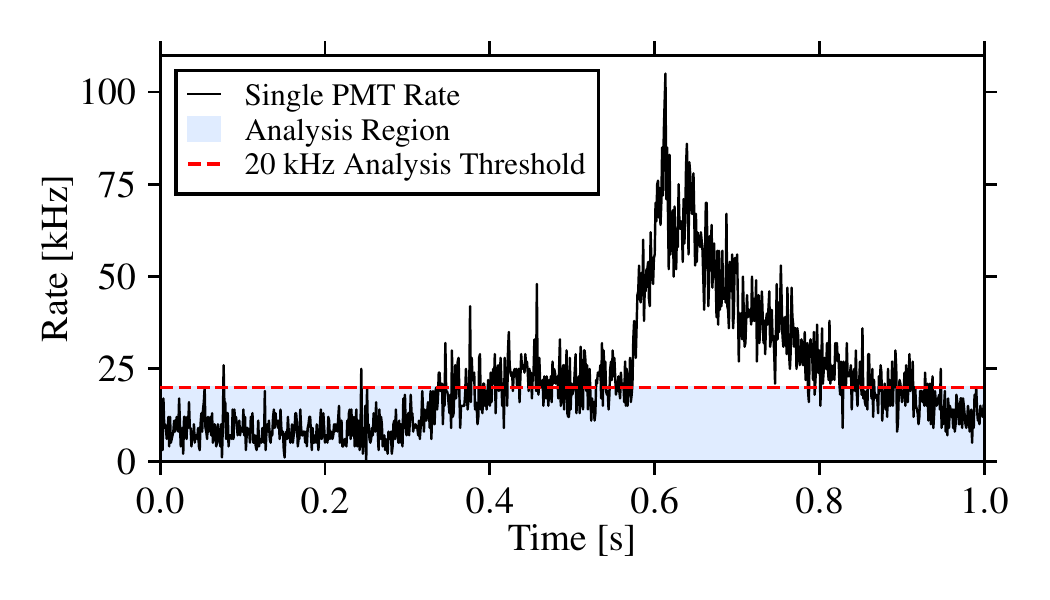}
\caption{One second of STRAW data showing the detection rate (black) of a single sDOM. The blue region represents the acceptable range of rates for the $^{40}$K analysis.}
\label{fig:STRAW-K40}
\end{figure}

Data
{\color{black}were}
taken in high-precision mode from all sDOMs and periods in which the rate was lower than $20\,\mr{kHz}$ were used. Fig. \ref{fig:STRAW-K40} shows one second of characteristic sDOM data with the sub-threshold data taking region used highlighted. In total, $16\,\mr{hours}$ of low activity data were analyzed for coincident events.

\subsection{Geant4 Simulation}

\begin{figure*}[ht]
\centering
\includegraphics[width=\textwidth]{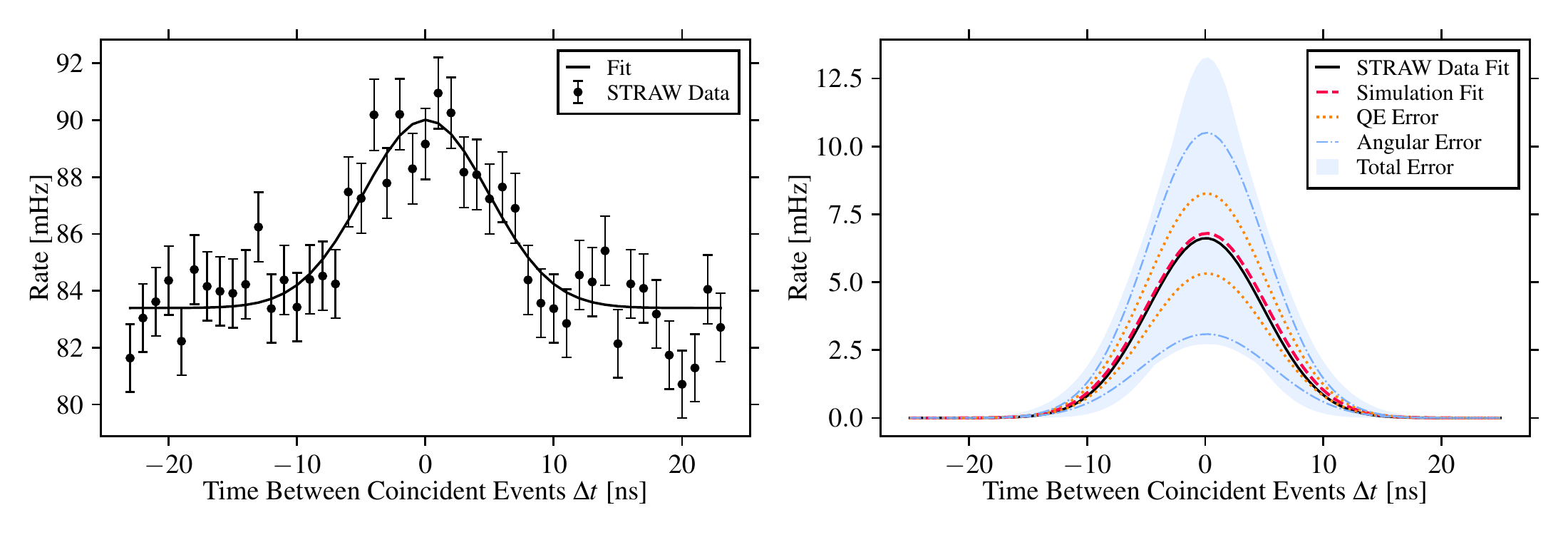}
\caption{\textbf{Left:} STRAW coincidence histogram with Gaussian fit. \textbf{Right:} Gaussian fits to coincident detection rate distributions of STRAW data with the baseline subtracted (black) and simulation (red) plotted with the total systematic error band (blue). The dotted and dashed-dotted lines represent bands corresponding to the error contributions from quantum efficiency (orange) and angular acceptance (blue). The individual error contribution from the absorption length is too small to be resolved in the plot so it is not shown.}
\label{fig:Results-K40}
\end{figure*}

A Geant4 \cite{Agostinelli2003} simulation was developed consisting of an sDOM at the origin of a $25\,\mr{m}$ radius sphere of seawater. Due to the back-to-back PMT geometry of the sDOMs, most coincident $^{40}$K photons arrive at large angles to the PMTs. At these large angles, the acceptance of the sDOM is low. The simulated seawater is characterized by the attenuation length obtained in Sec. \ref{sec:modelfit} along with the abundance of $^{40}$K estimated from the monitored salinity at Cascadia Basin.
{\color{black} Both decay channels, $\beta^-$ decay and electron capture (eqs. \ref{eq:betaminus} and \ref{eq:ecap}) are considered in the simulation.} Based on the total $^{40}$K activity, {\color{black} decay products} were randomly generated throughout the volume for the equivalent of {\color{black} $3.0\,\mr{minutes}$  with energies distributed according to the expected spectrum of the $^{40}$K $\beta^-$ and electron capture decays \cite{Fiorentini2007}.}
{\color{black}The optical transmittance of the glass, the quantum efficiency, and the transit time spread of the PMTs were the detector parameters included in the simulation.}
The mean transit time spread of $6.5\,\mr{ns}$ assumed in the simulation was taken from lab measurements of a PMT at large incidence angles.

Systematic errors associated with the sDOM geometry, quantum efficiency, absorption lengths, and variations in the transit time are included in the analysis. The first three are applied as relative uncertainties to each detection, whereas the variation in transit time is applied as an additional $\Delta t$ Gaussian smearing of $\pm1\,\mr{ns}$.

\subsection{Results}

Fig. \ref{fig:Results-K40} (left) shows the distribution of coincident detection rates for the STRAW data. As expected, a baseline of random events and a peak centred at $\Delta t = 0$ caused by coincident photons of the same $^{40}$K decay are visible. To compare the simulation results with STRAW data, the same type of distribution was generated for simulated data and fit with a Gaussian. Fig. \ref{fig:Results-K40} (right) shows a comparison between the two Gaussian fits along with the systematic uncertainties associated with the simulation. In this comparison, the baselines of both fits were removed because the simulation only accounts for $^{40}$K whereas the baseline in STRAW data has many other contributing factors that were not accounted for.

Fits to both simulation and data are in good agreement. Using this result, the accuracy of simulation inputs can be checked by comparing the ocean salinity determined from data and simulation to the salinity independently measured by Ocean Networks Canada. The coincident detection rate $R$ of $^{40}$K decays in a single sDOM can be written as the product of the $^{40}$K decay rate per unit volume, $B_q$, and the effective detector volume for  coincident $^{40}$K detection $V_{\text{eff}}$:
\begin{equation}
    R=B_q\cdot V_{\text{eff}}
    \label{eq:K40_Rate}
\end{equation}
$V_{\text{eff}}$ can be determined directly from the simulation as:
\begin{equation}
    V_{\text{eff}}=\dfrac{n_{\text{det}}}{n_{\text{gen}}}V_{\text{gen}}
    \label{eq:K40_Veff}
\end{equation}
Where $n_{\text{det}}$ is the number of detected coincidences, $n_{\text{gen}}$ is the number of generated electrons, and $V_{\text{gen}}$ is the generation volume. From simulated results, {\color{black} $V_{\text{eff}}=9.1\pm5.1\,\mr{cm^3}$.}

To determine $R$ from STRAW data, the integral of the Gaussian fit is used.
\begin{equation}
    R=\dfrac{a\sigma \;{\sqrt[]{2\pi}}}{\Delta \tau}
    \label{eq:K40_R}
\end{equation}
where $a$ and $\sigma$ are the amplitude and standard deviation of the Gaussian and $\Delta\tau$ is the bin width of the distribution. Using eq. \ref{eq:K40_Rate}, the $^{40}$K decay rate per unit volume is given by:
\begin{equation}
    B_{q}=\dfrac{1}{V_{\text{eff}}}\dfrac{a\sigma \;{\sqrt[]{2\pi}}}{\Delta \tau}
    \label{eq:K40_Bq}
\end{equation}
 This yields a $^{40}$K activity rate {\color{black} $B_q=8.6\pm4.7\,\mr{\tfrac{Decays}{ms~m^3}}$.} The activity rate is related to ocean salinity, $r_s$, according to:
\begin{equation}
    r_{s}=\dfrac{B_{q}\tau_{1/2}A}{r_{K}r_{I}\rho N_{\mr{A}}\ln{2}}
    \label{eq:K40_rs}
\end{equation}
Where $r_k$ is the potassium fraction in sea salt, $r_I$ is the isotope fraction of $^{40}$K, $\tau_{1/2}$ is the halflife of $^{40}$K, $N_{\mr{A}}$ is Avogadro’s number, and $A$ is the atomic weight of $^{40}$K \cite{Albert_2018}. A salinity of
{\color{black} $2.5\pm1.4$\%} is found, which covers the salinity measured by ONC of $3.482\pm0.001$\% {\color{black}\cite{oceans2}}.
This confirms the validity of the simulation and gives confidence that the water properties used as simulation inputs are correct.

\section{Bioluminescence and Background Rates}

Bioluminescence is the emission of light by living organisms triggered by either mechanical, electrical, or optical stimulation. In the marine environment, it is a pervasive mechanism used by a wide range of species, from bacteria to large fish, for finding food, attracting mates, and evading predators \cite{Widder704}. A recent study has shown that close to 75$\%$ of all organisms larger than 1$\,$cm living between the surface and a depth of 4000$\,$m are capable of bioluminescence \cite{Martini}.

While the bioluminescence emission spectrum varies greatly among species, the bulk of emission occurs in the blue spectrum, between 440-500$\,$nm, where the absorption length of water is highest \cite{Widder704}. A precise characterization of this phenomenon is necessary to quantify its impact on the telescope background levels, in particular as physical structures exposed to turbulent flows are known to trigger bioluminescence \cite{Widder2}. In this context, long-term monitoring of local and diffuse bioluminescence activities has been performed by STRAW.

\subsection{Method}

The STRAW low-precision data
{\color{black}were}
used to monitor the environmental conditions at the Cascadia Basin. The registered rates are mainly due to three factors:  the photomultiplier dark noise, the $^{40}$K radioactive decays, and the ambient bioluminescence. This combination can change over time according to different environmental conditions.

The data
{\color{black}are}
analyzed by looking at the range of rates for the lowest sDOM threshold, set at half the photo-electron level. Rates are measured over a $30\,\mr{ms}$ window (sec. \ref{setup}). The distribution of rates over two years is studied along with the variation of these rates over time. Since for a small fraction of time the rates exceed the DAQ capabilities, the variation over time is studied using percentiles, rather than mean values, as they remain unaffected by this.

\begin{figure}[ht]
\centering
\includegraphics[width=0.48\textwidth]{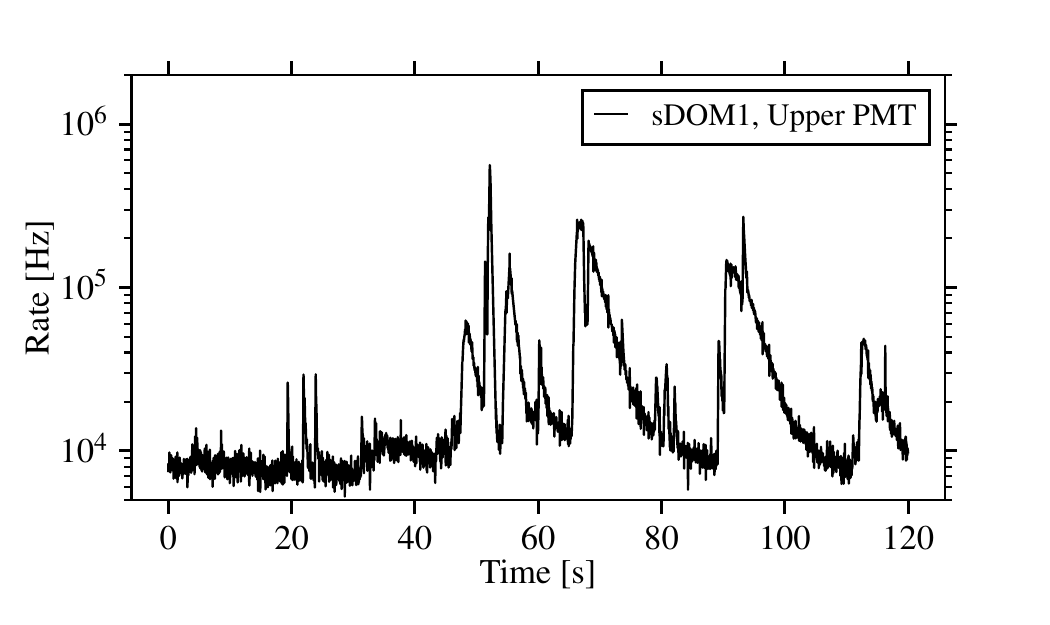}
\caption{Rate of a single photomultiplier over two minutes, measured in $30\,\mr{ms}$ intervals. There is a characteristic structure of a constant background with spikes caused by bioluminescence, typically
{\color{black}of}
the order of a few seconds.}
\label{fig.small_time_window}
\end{figure}

An example of two minutes of data for the upper PMT of sDOM1 is shown in fig. \ref{fig.small_time_window}. The rates follow the typical bioluminescence structure, with spikes on top of a constant background level, ranging from a few kHz to several MHz.

\subsection{Observations}

\begin{figure}[ht]
\centering
\includegraphics[width=0.48\textwidth]{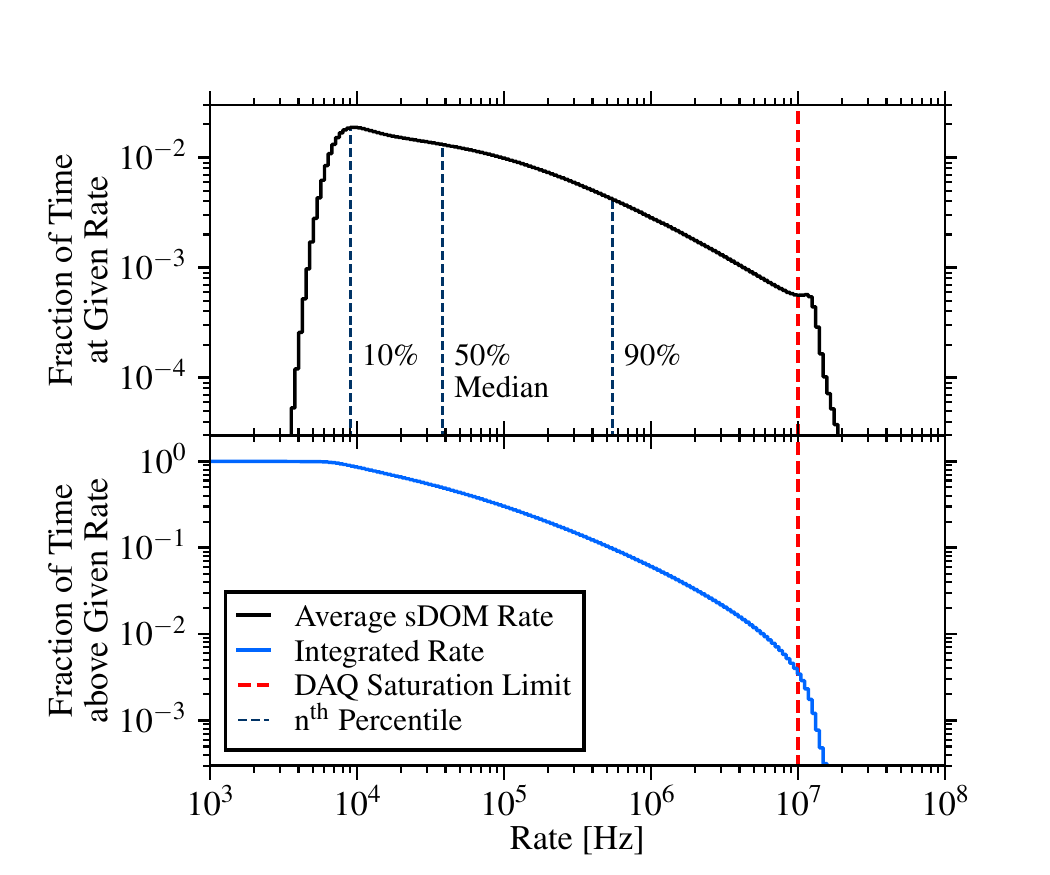}
\caption{Distribution of the single-PMT background rates measured over two years of STRAW operation.
The $10^\mr{th} {\color{black}\:(9\,\mr{kHz})}$, $50^\mr{th} {\color{black}\:(38\,\mr{kHz})}$, and $90^\mr{th} {\color{black}\:(548\,\mr{kHz})}$ percentile are shown,
their evolution over time is shown in figs. \ref{fig:four-days} and \ref{fig:two-years}. The bottom plot shows the integral fraction of time from a given rate to infinity, showing for which fraction of time a certain rate was exceeded.}
\label{fig:STRAW-distribution}
\end{figure}

The distribution of rates over two years is shown in fig. \ref{fig:STRAW-distribution} (upper plot). The lower limit corresponds to the $^{40}$K and dark noise baseline level, while the bioluminescence rates vary over several orders of magnitude, occasionally exceeding the maximum detection rate of $10\,\mr{MHz}$. This is a fundamental input for the design of the future P-ONE DAQ system. In addition, the fraction of time above a given rate has been computed (fig. \ref{fig:STRAW-distribution}, lower plot) which can be used to estimate the bioluminescence-induced dead-time of such a DAQ system.

\begin{figure}[ht]
\centering
\includegraphics[width=0.48\textwidth]{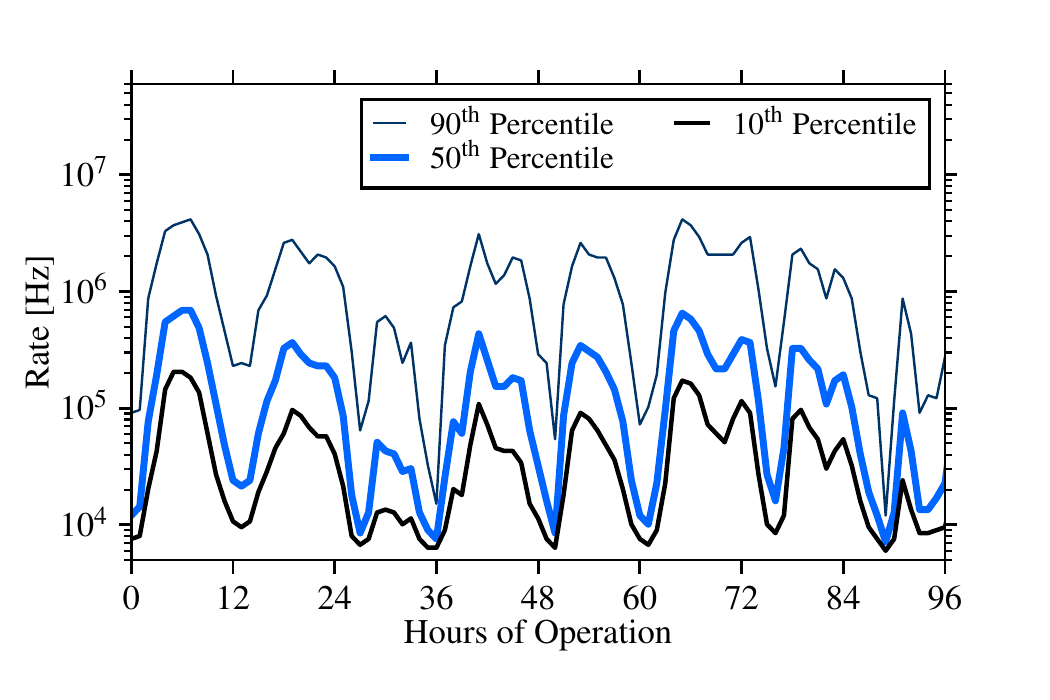}
\caption{Hourly percentiles of the background rates over four days. A clear time structure corresponding to the 12.5 hour tidal cycle is visible. The percentiles are chosen instead of the mean rate, as the percentiles are unaffected by the DAQ saturation limit (see fig. \ref{fig:STRAW-distribution}).}
\label{fig:four-days}
\end{figure}

\begin{figure}[ht]
\centering
\includegraphics[width=0.48\textwidth]{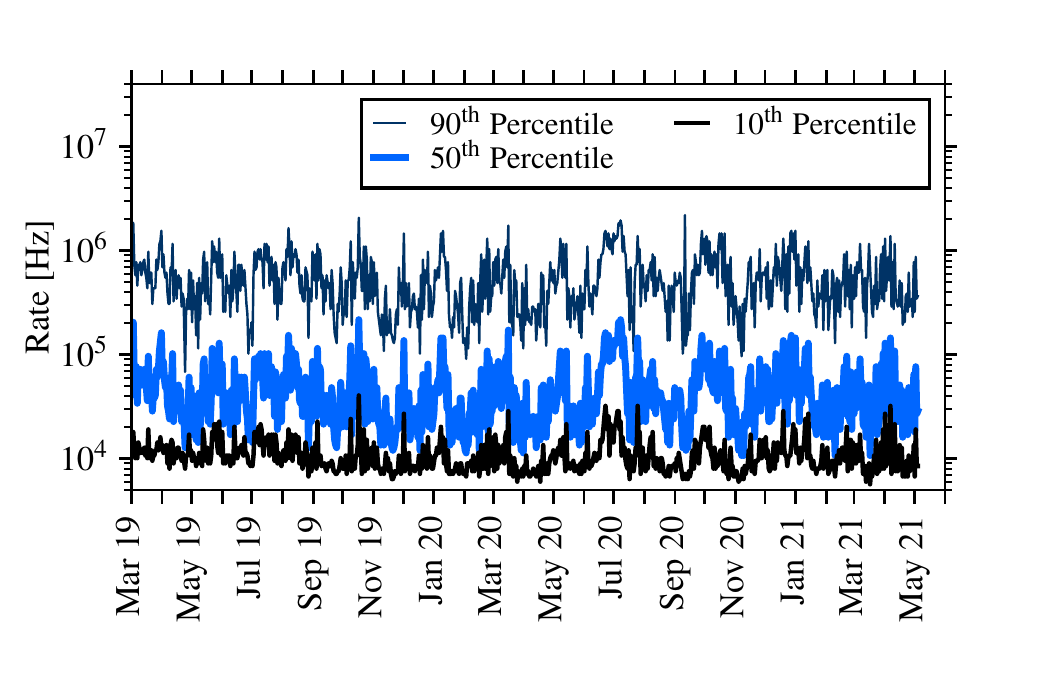}
\caption{Daily percentiles of the background rates over two years of STRAW operation. A detailed investigation of the long-term bioluminescence monitoring will be the subject of a future paper.}
\label{fig:two-years}
\end{figure}

To study the change of rates over time, the 10th, 50th (median), and 90th percentile have been calculated and their development over four days (fig. \ref{fig:four-days}) and over two years (fig. \ref{fig:two-years}) is shown. The percentiles were calculated on an hourly basis over four days and then on a daily baseis over the entire two-year time window. Fig. \ref{fig:four-days} shows a modulation of about 12.5 hours over the different percentile values. This value corresponds to the tidal cycle.
Fig. \ref{fig:two-years} shows the median rates changing between 10$\,$kHz and 100$\,$kHz over two years, with no significant seasonal variation.

It is important to note that the rates are specific to the 3" PMTs used in STRAW. For other PMT sizes, the rates will scale with the photocathode area.

\section{Conclusion}

The measurements of the attenuation length show that the Cascadia Basin is comparable to other water-based neutrino detector sites, with an attenuation length of $27.7\substack{+1.9\\-1.3}\,\mr{m}$ at $450\,\mr{nm}$
{\color{black} and falls within the requirements laid out in \cite{Agostini2020}}.
In addition, measured coincident event rates due to potassium-40 decays agree with the simulated predictions. The background rate, dominated by bioluminescence, reaches from $10\,\mr{kHz}$ to {\color{black}several} $\mr{MHz}$ ({90}th percentile), with strong variations over time.

\section{Outlook}
The Cascadia Basin has been optically characterized by the STRAW
{\color{black}pathfinder}
as a suitable site for the future P-ONE neutrino telescope. A
{\color{black}second pathfinder,}
called STRAW-b, has been deployed in October 2020 next to STRAW. STRAW-b is collecting data to extend and complement the results reported here.

Using the experience gained from STRAW, the design of the P-ONE neutrino telescope is now underway. The development of P-ONE will complement other experiments, moving closer to a global network of neutrino telescopes. Together, the telescopes in this network will cover almost the entire sky, increase the global observation rate of high-energy neutrinos and expand the bounds of neutrino astronomy.

\bibliography{pone}

\end{document}